\documentclass[11pt]{article} 
\usepackage[margin=1in]{geometry} 
\usepackage{setspace} 
\usepackage{graphicx}
\graphicspath{ {images/} }

\usepackage[autostyle, english = american]{csquotes}
\MakeOuterQuote{"}

\usepackage{dcolumn} 

\usepackage[hidelinks]{hyperref}
\usepackage[english]{babel}

\usepackage{parskip}
\usepackage{xcolor}

\usepackage{authblk}

\usepackage{amsmath}
\usepackage{cite}

\usepackage{pdflscape}

\usepackage{comment}

\usepackage{natbib}
\bibpunct[, ]{(}{)}{,}{a}{}{,}%
\def\bibfont{\small}%

\usepackage{endnotes}

\let\footnote=\endnote

\usepackage{natbib}
 \bibpunct[, ]{(}{)}{,}{a}{}{,}%
 \def\bibfont{\small}%

\usepackage{color}
\usepackage{url}
\usepackage{pbox}
\usepackage{amssymb}
\usepackage{amsmath}
\usepackage{hyperref}
\usepackage{breakurl}
\usepackage{graphicx}
\newcommand{\LeftEqNo}{\let\veqno\@@leqno}
\usepackage{environ}
\makeatletter
\NewEnviron{Lalign}{\tagsleft@true\begin{align}\BODY\end{align}}
\makeatother
\usepackage{xcolor}
\hypersetup{
	colorlinks,
	linkcolor={red!50!black},
	citecolor={blue!50!black},
	urlcolor={blue!80!black}
}
\usepackage{float}
\usepackage[caption = false]{subfig}
\usepackage{graphicx}
\usepackage{tikz}

\usetikzlibrary{arrows}
\usetikzlibrary{calc}
\usepackage{algorithm}
\usepackage[noend]{algpseudocode}
\usepackage{booktabs}

\newcommand{\bpDemand}{g}
\newcommand{\aoDemand}{a}
\newcommand{\distance}{d}
\newcommand{\distanceMatrix}{D}
\newcommand{\currentSites}{S}
\newcommand{\allSites}{N}
\newcommand{\candidateSizes}{Q} 
\newcommand{\expansionNumSites}{K}
\newcommand{\currentNumSites}{s}
\newcommand{\totalNumSites}{n}
\newcommand{\candidateNumSites}{q} 
\newcommand{\householdIncome}{h}
\newcommand{\population}{p}

\newcommand{\set}[1]{\left\{#1\right\}}

\newcommand{\card}[1]{\left|#1\right|}

\newcommand{\svrCost}{C}
\newcommand{\weightMatrix}{W}
\newcommand{\numberSupportVector}{m}
\newcommand{\tengcolor}[1]{\textcolor{black}{#1}}
\newcommand{\svrEpsilon}{\varepsilon}
\newcommand{\errorTerm}{\epsilon}
\newcommand{\linearCost}{l}

\newcommand{\supportVector}{V}

\newcolumntype{C}[1]{>{\centering\let\newline\\\arraybackslash\hspace{0pt}}m{#1}}

\usepackage{lscape}
\usepackage{afterpage}
\usepackage{longtable}

\usepackage{setspace}

\allowdisplaybreaks

\begin{document}
	
	\title{Predictive and Prescriptive Analytics for Location Selection of Add-on Retail Products} 
\author[1]{Teng Huang\thanks{Teng.Huang@uconn.edu}}
\author[1]{David Bergman\thanks{David.Bergman@uconn.edu}}
\author[1]{Ram.Gopal\thanks{Ram.Gopal@uconn.edu}}
\affil[1]{Department of Operations and Information Management, School of Business, University of Connecticut}
	\date{}
	\maketitle
	\providecommand{\keywords}[1]{\textbf{\textit{Keywords:}} #1}
	
	\abstract{
In this paper, we study an analytical approach to selecting expansion locations for retailers selling add-on products whose demand is derived from the demand of another base product. Demand for the add-on product is realized only as a supplement to the demand of the base product. In our context, either of the two products could be subject to spatial autocorrelation where demand at a given location is impacted by demand at other locations. Using data from an industrial partner selling add-on products, we build predictive models for understanding the derived demand of the add-on product and establish an optimization framework for automating expansion decisions to maximize expected sales. Interestingly, spatial autocorrelation and the complexity of the predictive model impact the complexity and the structure of the prescriptive optimization model. Our results indicate that the models formulated are highly effective in predicting add-on product sales, and that using the optimization framework built on the predictive model can result in substantial increases in expected sales over baseline policies.
}

		
		\newpage

\section{Introduction}
\label{sec:intro}

Managers of retail chains face constant pressure to increase their sales. Given that the geographic reach of one retail store is necessarily limited \citep{kumar1988measuring, pancras2012empirical}, managers often resort to the strategy of opening new outlets to achieve higher sales. However, the effect of opening new outlets on the performance of a whole retail chain is complex \citep{pancras2012empirical}. Picking the outlet that generates the largest expected sales on its own will not guarantee the largest marginal revenue of the whole retail chain.

Location is important to retailers, especially for those whose services have to be made through fixed geographic locations \citep{brown1981innovation,ketelaar2017disentangling}. Access to retail outlets is a critical factor in determining patronage \citep{ghosh1986approach,eze2015correlation}. A good location strategy also gives the retailer strategic advantages over its competitors \citep{jain1979evaluating, obeng2016survival}. The location decision usually represents a large-amount long-run fixed investment \citep{craig1984models,oner2016retail}. When considering adding more service locations, the retailer trades off between the expected revenue with the additional cost of providing the service \citep{ghosh1986approach, pancras2012empirical}. In other words, there shouldn't be expansion unless there is sufficient increase in profit. Moreover, the manager should not only consider a single separate location, but also the whole retail network.

This paper focuses on retail network expansion in the context of \emph{add-on products}. The specific products that we study in this paper are those for which:
\begin{enumerate}
	\item demand is realized only as a supplement to the demand for another product (which we refer to as the \emph{base product})
	\item purchases are made as optional additional purchases at the point of sale of base products
	\item base-product sales are done in standard brick-and-mortar (non-online) retail outlets.
\end{enumerate}
The add-on products we study are therefore of the following form. First, the demand is derived from the demand for another product. Second, the add-on product can only be sold together with the base product.  Third, the base product is sold at certain physical locations, and both products are purchased at these locations.

There are many real-world settings where products of this form are sold. One such setting is in \emph{car maintenance}.   For example:
\begin{enumerate}
	\item Fuel additives sold at-the-pump; the base product is gasoline, and the add-on product is the additive.
	\item Tire polish services sold at car wash places; the base service is a car wash, and the add-on product is the tire polish.
	\item Air filter replacement at oil change; the base product is the oil change, and the add-on product is the air filter.
\end{enumerate}
The company selling the add-on product that we partner with is an add-on retailer for a car maintenance product, and so we focus the discussion on this context.  Note that the setting is not limited to car maintenance, where similar examples are found ubiquitously in many industries, for example with body care products sold during personal care services and private training lessons in fitness clubs, for example.

 The problem we study is how to maximize expected sales of add-on products by expansion, where a company can only expand to a limited number of locations among those where the base product is currently offered.  We present a decision-making framework for determining the optimal location for expansion.  This requires accurate predictive models of demand given a set of base product retail locations, and an optimization framework for automating the selection of an optimal collection of expansion sites. 
 
For predicting sales, we apply several machine learning algorithms and evaluate their performance on a year of point-on-sales data at existing sites where add-on products are sold.  We use basic area-specific demographic information (median household income and population) to supplement base-product and add-on product demand data to create more accurate models.  
 
 We further supplement the data set with a spatial-weight matrix, whose entries provide a measure of geographic closeness between base-product retail locations that decays as the distance grows.  Including this matrix as part of an input to the learning models is shown to increase predictive accuracy only when spatial autocorrelation exists in base-product demand, as exhibited through an experimental evaluation on two different regions, one in the presence of and the other in the absence of spatial autocorrelation, as indicated by the \emph{Moran's $I$ Test} \citep{moran1950notes, bivand2008applied}.
 
 The models we develop in this paper have predictive power better than other models suggested in the literature for similar contexts; for example, the (out-of-sample) \emph{mean absolute percent error} of our predictive models is around 20\%, which is less than that reported by \cite{fisher2016optimal}, for predicting sales at online websites.  
 
 Equipped with an accurate predictive model for sales in a region, we then turn to formulation and solving an optimization model for automating expansion decision making.  The optimization model can take various forms, depending on the predictive model employed.  We describe in detail the form the optimization model takes for the various predictive models tested, and suggest a simple general-purpose heuristic that can be used for any such model.   
 
 In comparison with a standard baseline policy for expansion, in which the add-on product company expands based on higher base-product sales, the predictive-and-optimization modeling framework developed can result in over 5\% additional sales, depending on the base-product demand distribution, the number of expansion sites that the add-on product company can afford, and the presence of spatial autocorrelation, simply by choosing better expansion locations.  The solutions obtained are shown to be robust, in that they result in favorable expansion decisions, even when evaluated on different predictive models than the one used in the optimization model.

The contributions of this research and the managerial insights they reveal are therefore as follows:
\begin{itemize}
	\item the design of accurate predictive models for the demand of add-on products using limited data;
	\item an understanding of when spatial-weight matrices can improve predictive accuracy based on spatial autocorrelation;
	\item an optimization framework for automating expansion decisions based on predictive models; and
	\item an application to a real-world data set where the predictive-and-optimization modeling framework provides substantially better solutions than baseline expansion policies. 
\end{itemize}

The remainder of the paper is organized as follows.  We first review related literature in Section~\ref{sec:letReview}. In Section~\ref{sec:ProblemFormulation} we provide details of the problem setting and context.  In Section~\ref{sec:data} we describe the data set provided by our industrial partner, and the various data elements that we gathered from publicly available sources. Section~\ref{sec:predictiveModels} describes the predictive models we employ and analyzes their accuracy on the data.  Section~\ref{sec:ExpansionOptimization} presents the optimization models, both in general and for specific predictive models, suggests a general-purpose heuristic for solving the optimization models, and reports on the effectiveness of the solution obtained. We conclude in Section~\ref{sec:conclusion}.

\section{Literature Review}
\label{sec:letReview}

This work is closely related to a variety of research streams.  We discuss each, highlighting the most relevant papers found in the literature to the problem we study.

\subsection{Machine Learning for Predicting Demand and Optimizing Decision Making}

In recent years, there have been a few studies focused on predicting demand for products using advanced machine learning algorithms. \cite{ferreira2015analytics} test linear regression models and regression trees for \tengcolor{the daily sales of an online retailer}, where they find that regression trees provide the best prediction. \cite{cui2017operational} implement multiple machine learning methods to forecast daily sales for an online apparel retailer. \cite{fisher2016optimal} propose a combined method to \tengcolor{predict demand at potential locations, which utilizes both machine learning methods and econometric techniques. }

Our work differentiates and builds upon this literature by studying add-on products, estimating their demand using machine learning algorithms, and then using optimization to determine best operational decisions.   We also introduce the idea of performing a robustness check, where one takes the solutions obtained by a predictive-and-prescriptive model and evaluate its performance on other predictive models. This ensures that the solutions obtained are robust and can be trusted by the organization employing them for decision making.

\subsection{Add-on Products}

There is a limited stream of research focused on understanding how add-on products affect base-product sales.  
This research includes studies on how add-on products provide vertical differentiation \citep{lin2017add}, and how they affect consumers' evaluation of a base product \citep{bertini2008impact}.  As an up-selling strategy, there is conflicting research on whether add-on products  harm consumer welfare \citep{gerstner1990can, wilkie1998does, wilkie1998doesadv, hess1998yes}. Other studies related to add-on products include pricing strategies for promotional products under up-selling \citep{aydin2008pricing} and pricing strategies for firms selling both base products and add-ons \citep{erat2012consumer}. 

To the best knowledge of the authors, this paper provides the first study focused on demand estimation and retail expansion for add-on products.  We furthermore distinguish from the literature, as we do not presuppose that the base product and the add-on product belong to the same firm.

\subsection{Retail Site Decision Making}

Location decision making in the retail industry is critical. There exists considerable literature dealing with retail location models \citep{craig1984models,Drezner201551}. The research most closely related to the work in this paper is by \cite{ghosh1986approach}, where \tengcolor{the authors design a network of service centers via choosing the network size, the location of outlets, and operating characteristics simultaneously}.  A major difference between this work and the present paper is that the authors do not consider competition among outlets. 
In addition, \cite{ghosh1986approach} deals with determining an optimal network design, including both the optimal size of the network given an exogenous demand and the characteristics that will attract customers. In our setting, the network is fixed, in that expansion decision for the add-on product is restricted to the locations that sell the base product.  

There is an increasing number of papers that consider the cannibalization effect among outlets of the same retailer \citep{nishida2014estimating, pancras2012empirical, fisher2016optimal}. \cite{pancras2012empirical} examines the impact of opening/closing an outlet on overall chain performance, but their demand model is more suitable for isolating locations that can be closed. 
Our focus is narrowed to expansion decision. Due to the add-on nature of our focal product, we operate under the assumption that the cost of keeping operation is lower than that of ending the contract with the institutes that operate the base product. Second, like most research on this topic, they estimate the parameters using panel data from a large U.S. city. This causes context dependency. Results from one study area cannot be expected to hold in other areas. This limitation restricts the use of choice models calibrated from one study region to design location strategies in another.

\subsection{Facility Location Problem}

The literature on facility location problems dates back at least one-hundred years \citep{WebIndTub1909}, with perhaps an even longer history dating back to studies of Pierre de Fermat, Evagelistica Torricelli (a student of Galileo), and Battista Cavallieri \citep{DreHam02}.  There are many books dedicated to the subject (e.g., \citealp{FraWhi74,HanMir79,LovMorWes88,Ros94,DreHam02,NicPue07,ChuMur09,Wol11,Das13}), as well as countless surveys and articles beyond this. Some authors have attempted to develop a taxonomy for facility location problem \citep{HamNicSch96,Das08} where models are broken down based on continuous/discrete/network demand and location decision, the number of facilities, the type of facilities, capacity restrictions, and objective functions.

The stream of literature most-closely related to the problem studied in this paper is the \emph{competitive facility location} (CFL) model  where facilities set up by decision makers will compete for market share and profitability. Hotelling \citep{Hot29} was perhaps the first to study location models with competition where he considers the case of two ice cream shops picking locations for their respective storefronts and customers choose a shop based on their relative distances. There have been a number of papers dedicated to classifying CFL models \citep{Dre92,EisLapThi93,Kres12,Kar15}.  The taxonomy is typically based on the spatial representation (continuous, discrete, network), distance measure (Manhattan distance, $\ell-2$ norm, etc.,), the nature of the competition (static, dynamic, sequential), the number of new locations (single versus multiple), the nature of the customers (demand is elastic versus inelastic, customers come from a single location or multiple locations, customers are served by a single location or by any number of locations based on a probability distribution, etc.), the objective type (customers prefer facilities nearby or far way), among other defining characteristics.

CFL models using spatial interaction typically employ a demand function for each customer, and that model is based on the probability that a customer will choose a particular facility based on the customer's proximity to a facility.  In our model, customer demand for our focal products is a function of the base-product sales and the sales in nearby locations (scaled by their proximity).  

Studies do appear in the literature that investigates CFL with cannibalization and market expansion. \cite{Boz10} study CFL with location routing where a firm incurs costs due to vehicles having to service each location from a central warehouse.  Other authors have investigated spatial interaction models using concave demand models \citep{AboBerKra07,AboBerKra07a,Aboolian2009} but again do not consider demand as derived from other products.  More recently there have been papers investigating models using a leader-follow framework, where one firm decides where to locate facilities and the other firm follows with a decision \citep{Drezner201551}. 

The problem studied in this paper differs from those appearing in the literature in that we study the demand of our focal products as a derived demand for another product.  

\section{Problem Context}
\label{sec:ProblemFormulation}

The goal of this paper is to develop a hybrid predictive-and-optimization modeling framework for automating the retail expansion location decision-making process for companies selling add-on products where base products are sold in fixed geographic locations.
We, therefore, seek to develop predictive models specifically designed for derived-demand products, and identify expansion locations based on those predictive models. We first develop a relationship between the profitability of retail outlets specifically for our focal product and the characteristics of their locations, and then we optimize the profit of the entire network.

We learn the relationship between the performance and the location characteristics from studying the past performance of existing locations. When retail managers make location expansion decisions, the attractiveness level of a location includes the demand, the buying power, and the competition level are common factors to consider. In addition, the demand for our focal product is derived from the demand for the base product. Hence, in the predictive models of the focal product's sales, we use the base-product sales, the weighted base-product sales in the neighborhood, the local median household income, and the local population as our predictive variables.

Before proceeding with the models, we fix notation.  Let $\allSites = \set{1, \ldots, \totalNumSites}$ be the set of sites selling the base product.  $\allSites$ is partitioned into two sets, $\currentSites$ and $\candidateSizes$, representing the sites currently selling the add-on product (\emph{active} sites) and the possible sites for expansion (\emph{candidate} sites), respectively.  Denote  by $\totalNumSites := \card{\allSites}, \currentNumSites := \card{\currentSites}$, and $\candidateNumSites := \card{\candidateSizes}$ the cardinality of the sets. Let $\bpDemand_i$ be the total sales of the base product at site $i$.  Furthermore, let $\expansionNumSites$ be the (maximum) number of expansion sites that the firm can choose.  We assume that expanding operations to any site requires a fixed cost which does not vary from site-to-site, and so, given a fixed budget for expansion, the firm can choose any set of sites within the allowable budget, or through management choice, to expand to, although this assumption can be relaxed.  

In our setting, an add-on product is offered for purchase each time a customer purchases the base product.  For any site $i \in \allSites$, let $\bpDemand_i$ be the number of times the base product is purchased at site $i$ (i.e., demand for the base product at size $i$).  In order to build accurate models for predicting purchases, we use demographic information common to the field studied.  In particular, we incorporate median household income $\householdIncome_i$ and zip-code population $\population_i$.  For notational convenience, we drop the index on parameters to represent the entire vector of values of that parameter (e.g., $\bpDemand \in \mathbb{R}^{\totalNumSites}$ is the $\totalNumSites$-dimensional vector of demands for all sites).

The first step in the framework is to develop a robust and accurate predictive model for the sales of the add-on product, $\aoDemand$. More formally, for any $\tilde{\allSites} \subseteq \allSites$, let $f(\tilde{\allSites})$
be the amount of the add-on-product sales given the firm's selection to offer the add-on product at locations $\tilde{\allSites}$.  We seek a model $\hat{f}$ that approximates $f$, using known data, which includes base-product sales at each location and median household income and population in the area of each location. 


Our models allow for the incorporation of location effects. In particular, the data we tested suggests that even for the same add-on product, that location effects can exist in particular regions or not. This can be formally tested through checking for \emph{spatial autocorrelation} through the Moran's $I$ test.  

After determining the best predictive model, we formulate a general optimization model for finding the best expansion sites. Namely, we solve:
\begin{equation}
\tag{EXP}
\label{eqn:optModel}
\begin{aligned}
&& \max_{\tilde{\allSites} \subseteq \allSites} \set{ \hat{f}(\tilde{\allSites}) : \textnormal{subject to } \tilde{\allSites} \supset \candidateSizes, |\tilde{\allSites}| = |\tilde{\currentSites}| +  \expansionNumSites.}
\end{aligned} 
\end{equation}

The complexity of this optimization problem depends on the underlying structure of $\hat{f}$.  We discuss the various forms this model takes based on the predictive model employed in Section~\ref{sec:ExpansionOptimization}.   We also devise a simple greedy heuristics for solving~(\ref{eqn:optModel}) given any functional form, which, despite its simplicity, works well and finds optimal solutions for all cases tested. For the problems in which we can find optimal solutions, the predictive models are of a form that when formulating~(\ref{eqn:optModel}), the problem reduces to a simple linear or quadratic binary optimization problem.

\section{Data}
\label{sec:data}

In order to evaluate the efficiency of the framework we develop, we use data provided by an add-on product retailer.  The retailer operates in over 1500  locations in the US, and partners with seven base-product companies, operating in locations that they currently do and can in the future sell their products.  The specific product is an add-on product in the context of car maintenance, that is sold at locations where customers of the base-product frequent and are offered the product on site at the time of purchase. 

The data utilized is at-the-site transactional data, that indicates each time a customer decides to purchase the add-on product when purchasing the base product.  This option is available each time the customer purchases.

\begin{figure}[t!] 
\centering
\subfloat{%
  \includegraphics[clip,width=0.6\columnwidth]{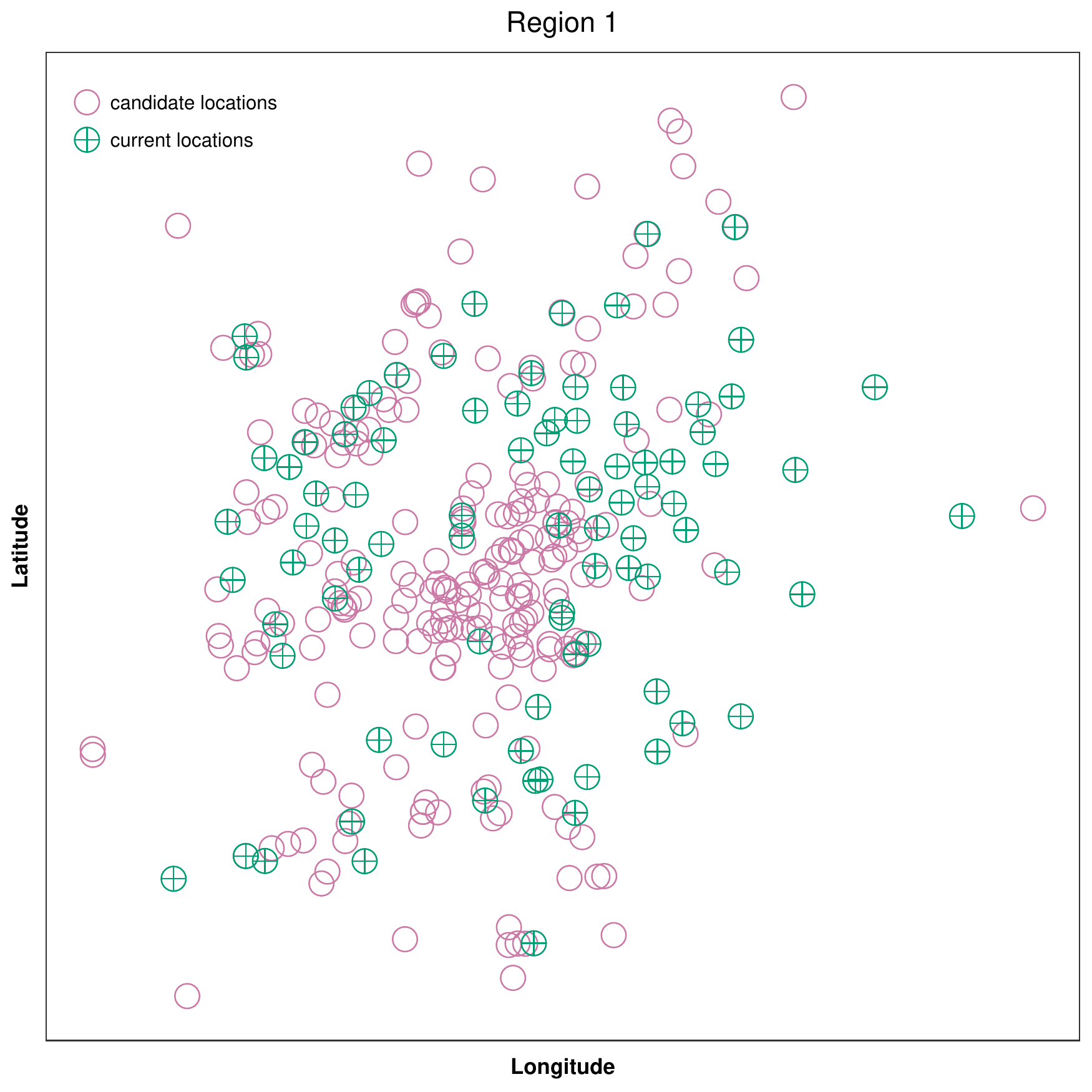}%
}

\subfloat{%
  \includegraphics[clip,width=0.6\columnwidth]{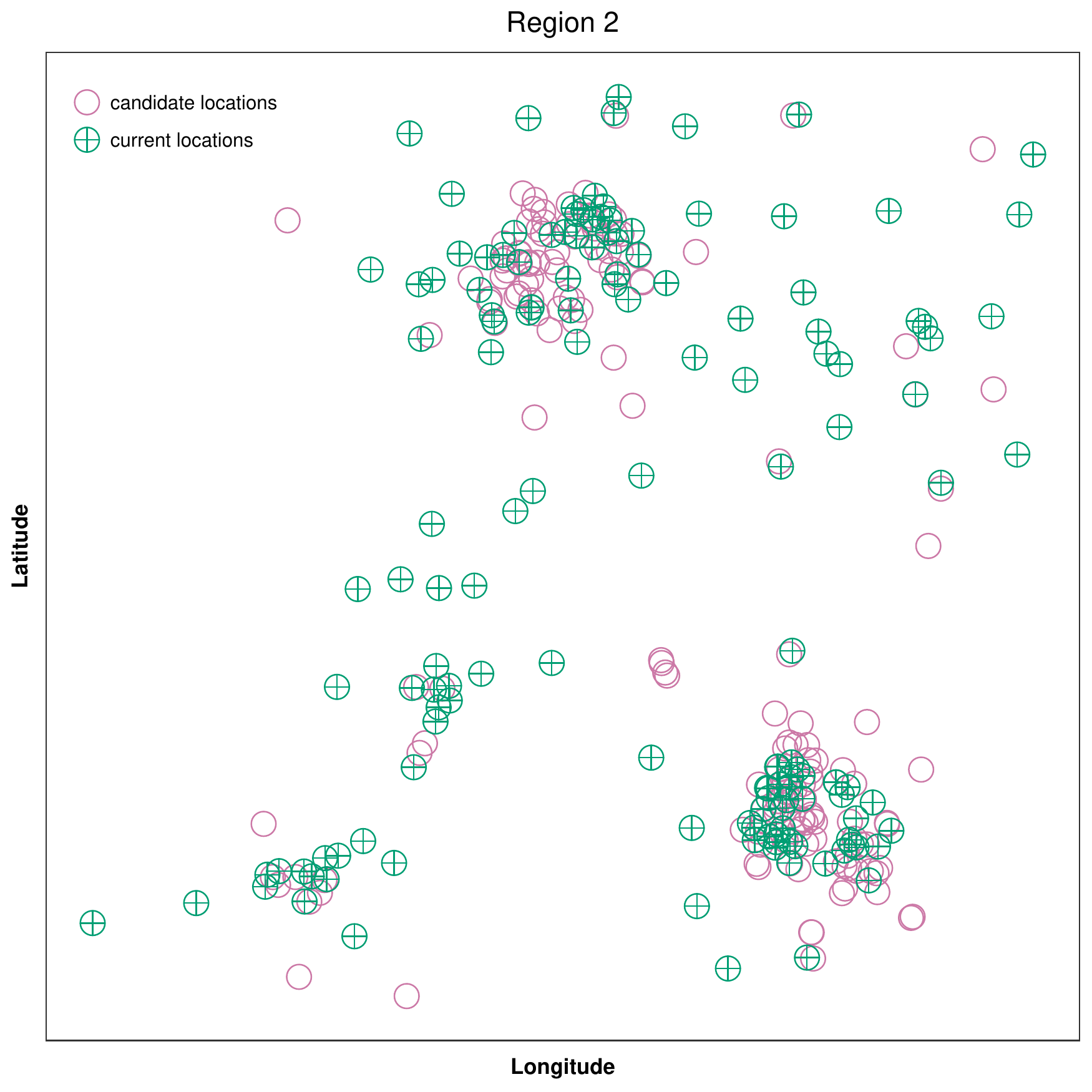}%
}

	\caption{Spatial distribution of base-product retail sites in  Region 1 and  Region 2.} 
	\label{fig:spatialDist} 	
\end{figure}

We select two geographical regions in the U.S. to study. Region 1 covers about 11878.59 $sq\ mi$, and has 89 existing operation sites and 228 candidate sites. The average distance between locations in this area is 17.2 miles. The locations are closely gathered as one cluster. Region 2 covers about 132549.2 $sq\ mi$, and has 146 existing operation sites and 162 candidate sites. The average distance between locations in this area is 54.8 miles. There are three clusters in this area. Figure~\ref{fig:spatialDist} depicts these regions. In both figures, the circles with crosses represent current locations, i.e., locations in $\currentSites$. The hollow circles represent candidate locations, i.e., locations in $\candidateSizes$. \tengcolor{We can see from Figure~\ref{fig:spatialDist} that there are relatively more expansion choices in Region 1 compared to Region 2. In both regions, the candidate sites are not evenly located. Some of them are more closely clustered, while some are far from any center. Our expansion strategy works well for both regions as demonstrated in Section~\ref{sec:ExpansionOptimization}.}

For each site $i \in \currentSites$, we have the aggregated point-of-sale transaction records of the outlets during 2015. 
We have the transaction records of both the base product and the add-on product. In addition, we have their geographic location (the latitude and the longitude in degrees), and the demographics, i.e., the median income per household and the population in a corresponding zip code. 
In summary, for each outlet, we have the number of the base product annual transactions, the number of the focal product annual transactions, its latitude and its longitude in degrees, the median income per household and the population in the corresponding zip code. 

The candidate locations $\candidateSizes$ we consider are those locations run by the base-product retailer where the add-on product is not currently sold. By the same mechanism used to gather data about current add-on sites, we have, for each location $i \in \candidateSizes$, $\householdIncome_i$ and $\population_i$.  The data from our partner company only lists base-product sales for the sites that the add-on product is currently sold at.   This provides us an opportunity to test different base-product demand profiles, in order to understand when the models we propose are particularly effective.   We discuss this in Section~\ref{sec:ExpansionOptimization}.

Summary statistics of the data is provided in Table~\ref{table:dataSummary}.  For Region 1 and Region 2, for $\currentSites$ and $\candidateSizes$, the data reports the mean and standard deviation of the base-product sales $\bpDemand$, the add-on product sales $\aoDemand$, the pair-wise distance between locations $\distanceMatrix$ (in miles), the household income $\householdIncome$ and the population $\population$. 

\begin{table}[h!]
		\begin{center}
			\begin{scriptsize}
			\caption{Summary statistics of industry partner data} 
			\label{table:dataSummary}
			\begin{tabular}{C{5ex}  C{12ex} | C{10ex} | C{10ex} | C{10ex}  C{0.5ex} C{12ex}  | C{10ex} | C{10ex} | C{10ex}  }
				& \multicolumn{4}{c}{Region 1} & & \multicolumn{4}{c}{Region 2}  \\ 
				\cmidrule{2-5}
				\cmidrule{7-10}
				& \multicolumn{2}{c}{$\currentSites$} & \multicolumn{2}{c}{$\candidateSizes$}
				&
				& \multicolumn{2}{c}{$\currentSites$} & \multicolumn{2}{c}{$\candidateSizes$} \\
				& Mean & SD &  Mean & SD & 
					& Mean & SD &  Mean & SD \\
					\cmidrule{2-5}
					\cmidrule{7-10} 
				 $\bpDemand$ & 196700.20& 82871.41& \textemdash & \textemdash & &195246.30& 93927.96& \textemdash & \textemdash  \\
				 $\aoDemand$ & 3420.58& 1149.45& \textemdash & \textemdash & &3087.76& 1548.83& \textemdash & \textemdash \\
				 $\distanceMatrix$ & 34.67& 18.21& 31.37& 18.30& &163.81& 88.08& 150.61& 98.03 \\
				 $\householdIncome$ & 67240.60& 18033.64& 62505.84& 11890.57& &62434.31& 22703.08& 69004.53& 18194.45 \\
				 $\population$ & 50641.40& 17718.66& 45898.06& 12815.12& &47208.74& 24986.17& 53736.55& 15935.17 \\
				 
			\end{tabular}
		\end{scriptsize}
		\end{center}
\end{table}

\section{Predictive Models}
\label{sec:predictiveModels}
	
This section provides details of the predictive models applied which will be used to optimize expansion decisions.

\subsection{Testing for Spatial Autocorrelation}

Before building the predictive models, we first look into the spatial interaction of both base-product sales and add-on-product sales, in the current sites. \tengcolor{First, georeferenced observations, for example, the two product sales at locations in vicinity, generally are not independent of one another \citep{getis2008history}. Strong evidence of the influence of spatial autocorrelation in the consumption of gasoline has been found \citep{dos2014spatial}.} Second, whether or not there is an impact on sales (either positive or negative) of either based on proximity can impact the best choice of predictive model.    We find that the best predictive model for a given region depends on whether or not spatial autocorrelation in the base-product sales exists.  As we will show, using or omitting a feature representing the total base-product sales scaled by the distance from the focal location has a positive impact on predictive models in the presence of spatial autocorrelation, and a negative impact in its absence.

In order to test whether there exists spatial autocorrelation in base-product sales, we apply a \emph{Moran's I} test.  This test is specifically designed to test whether or not spatial autocorrelation exists. 
The test statistics in this test, $I_g$, is calculated as a ratio of the product of the variable of interest and its spatial lag, with the cross-product of the variable of interest, and adjusted for the spatial weights used:
\[
I_{\bpDemand}
= 
\frac{\currentNumSites}{\sum_{i \in \currentSites} \sum_{j \in \currentSites} w_{i,j}}    
\cdot
\frac{\sum_{i \in \currentSites} \sum_{j \in \currentSites} w_{i,j} (\bpDemand_i - \bar{\bpDemand}) (\bpDemand_j - \bar{\bpDemand})}{\sum_{i \in \currentSites} (\bpDemand_i - \bar{\bpDemand})^2} \textnormal{,} 
\]
where $\bar{\bpDemand}$ is the average base-product sales over sites in $\currentSites$. The test assumes, in the absence of spatial autocorrelation, that this statistic follows a normal distribution. 
\tengcolor{In other words, if the test result is not significant, we cannot reject the null hypothesis that the observations are random.}
We discuss the results of applying the test to the sites in both regions in Section~\ref{sec:analysisPM}.

The feature we include to encode the scaled base-product sales is an inverse distance matrix $\weightMatrix$ times $\bpDemand$, the vector of gas sales.  In particular, for any two sites $i,j \in \allSites$, define $w_{i,j} := \frac{1}{\distance_{i,j}}$, where $\distance_{i,j}$ is the geographic distance (taken as the Eucledian distance) between the two sites.  In this way, $\weightMatrix \bpDemand$ is a vector that in coordinate $i$ contains the value $\sum_{j \in \tilde{\allSites}, j \neq i} \frac{1}{\distance_{i,j}} \bpDemand_j$.

\subsection{Demand Prediction Models}

In order to create the most accurate predictive models for estimating base-product sales, 
we apply linear regression models as a parsimonious baseline, and \emph{support vector regression} (SVR) models as a more advanced and nonparametric method for predicting expected add-on product sales. Our selection of predictive models was done for the following reasons.  Linear regressions is a classical tool that has been used ubiquitously throughout the literature, and provide a good baseline.  SVRs are used because they provide another more advanced mechanism for predicting sales, which allows input data to be transformed via \emph{kernel} functions in order to accommodate nonlinearity.   Our choice of prediction model among other available options (e.g., random forests, neural networks) is that SVRs provide closed-form solutions (which is critical for optimal expansion decision making), and that in other papers estimating demand using predictive models, SVRs provide among the best predictive accuracy \citep{cui2017operational, fisher2016optimal}.
\tengcolor{Random forests are shown to provide outstanding predictions \citep{cui2017operational, fisher2016optimal}. However, random forests do not work well in our spatial setting with a limited set of features used for prediction. Random forests rely on resampling both features and data in order to build prediction models. Features we employ cannot be expected to offer significantly more information when being resampled in building the random forests. Second, the construction of $\weightMatrix$ relies on the entire training data. When data is resampled with replacement, it is unreasonable to have two sites at the same location. Consequently, random forests would be nearly identical to a single decision tree if we resample without replacement. Therefore, we employ SVR models and linear regression models for prediction.}
In our preliminary tests, the linear-kernel and radial-kernel SVR models outperform those with a sigmoid kernel or a polynomial kernel, and therefore we only apply these for the remainder of this paper.

Our goal is to predict the add-on product sales given any set of sites.
In our derived demand scenario, the demand for the add-on product is derived from the demand for the base product. Hence, a portion of the predicted add-on product sales comes from sales of the base product.
In addition, the probability that a customer patronizes a facility is proportional to the attractiveness level and to a distance decay function \citep{berman2009modeling} and the attractiveness of a location is related to its demographics.
We include the base-product sales, the local median income per household and the total population as the independent variables, along with $\weightMatrix \bpDemand$.

\tengcolor{Price might be a strong indicator for sales. However, in our case, the price of the add-on product is fixed across the year we study. The price of the base product doesn't vary much across locations, especially after we take the average of the entire year. The mean of the base-product price in Region 1 is 2.21, while its standard deviation is 0.076. The mean of the base-product price in Region 2 is 2.09, while its standard deviation is 0.071. Hence, we don't include the price in our prediction models.}

In summary, the predictive models we apply are as follows (In regions without the spatial autocorrelated base-product sales, we remove $\weightMatrix \bpDemand$.):
	\begin{align*}
	& \textnormal{Linear Regression:} && \aoDemand = \beta_1 \bpDemand + \beta_2 \weightMatrix \bpDemand + \beta_3 \householdIncome + \beta_4 \population + \beta_0 + \errorTerm \\
	& \textnormal{Linear-Kernel SVR:} && \aoDemand = \omega \cdot \mathrm{ker}_{LK}(\bpDemand , \weightMatrix \bpDemand , \householdIncome, \population) + b_0 + \errorTerm \\
	& \textnormal{Radial-Kernel SVR:} && \aoDemand = \omega \cdot \mathrm{ker}_{RK}(\bpDemand , \weightMatrix \bpDemand , \householdIncome, \population) + b_0 + \errorTerm 
	\end{align*}

where,
\begin{itemize}
	\item $\aoDemand$ is the add-on product sales;
	\item $\bpDemand$ is the base-product sales;
	\item $\weightMatrix \bpDemand$ is the spatially weighted average of the base-product sales;
	\item $\householdIncome$ is the local median household income;
	\item $\population$ is the local population;
	\item $\omega$ is the estimated weights in the SVR models;
	\item $\mathrm{ker}_{LK}$ is the linear kernel function, i.e., $(x, x')$;
	\item $\mathrm{ker}_{RK}$ is the radial kernel function, i.e., $exp(-\gamma||x-x'||^2)$;
	\item $b_0$ is the estimated constant term from the SVR models; and
	\item $\errorTerm$ is the error.
\end{itemize}

\subsection{Analysis}
\label{sec:analysisPM}

We begin with the results from the Moran's $I$ tests, obtained using the \texttt{moran.test} (for the sales of the two products) and \texttt{lm.morantest} (for the residuals of a linear regression model of the add-on-product sales on the base-product sales) (in package \texttt{spdep} in \texttt{R}, version \texttt{R.3.4.2}) \citep{bivand2008applied}.  
Table~\ref{table:moran181} reports the $p$-values of the Moran's $I$ tests for the base-product sales, the add-on-product sales, and the residuals, in both regions of interest. 
Neither tests of the two product sales result in statistically significant spatial autocorrelation for Region 1, while with the opposite true for Region 2.   
We, therefore, conclude the following:
\begin{itemize}
	\item Neither the base-product sales nor the add-on-product sales in Region 1 are spatially autocorrelated. 
	\item Both the base-product sales and the add-on-product sales in Region 2 show spatial autocorrelation. In addition, after controlling for base-product sales, the residuals of the add-on-product sales show insignificant spatial autocorrelation, allowing us to conclude that the spatial autocorrelation in the focal product sales is inherited from the spatial autocorrelation of the base-product sales. 
\end{itemize}

\begin{table}[!htbp]
	\caption{Results ($p$-values) from Moran's $I$ test}
	\begin{center}
		\begin{tabular}{l | l | l | l}
			\hline
			Region & Add-on-product sales & Base-product sales & Residuals \\
			\hline
			Region 1 & 0.2568 & 0.1986 & 0.6506 \\
			Region 2 & 0.02739** & 0.05601* &  0.6291\\
			\hline
		\end{tabular}
	\end{center}
	\label{table:moran181}
\end{table}

\tengcolor{Generally speaking, if the Moran's I test statistic is statistically significant and positive, then the spatial distribution of high/low values is more spatially clustered than would be expected if underlying spatial processes were random \citep{ArcGIS}. Based on the test results in Table~\ref{table:moran181}, this is the case for Region 2. We can observe three clusters on Figure~\ref{fig:spatialDist}. On the other hand, for Region 1, we cannot reject the possibility that the spatial distribution of feature values is the result of random spatial processes.}

We conduct grid search over 50-time 10-fold cross validation to tune the hyper-parameters in the two SVR models. For the linear-kernel SVR model, we need to tune the cost $\svrCost$ and $\svrEpsilon$. For the radial-kernel SVR model, we need to tune $\svrCost$, $\gamma$, and $\svrEpsilon$. We compute the \emph{root mean squared error} (RMSE) as a measurement of how the combination of the hyper-parameter values performs. We pick the setting of the hyper-parameter values with the lowest RMSE. We start search with $\svrCost$ in $\set{2^0, 2^1 \ldots, 2^{16}}$, $\svrEpsilon \in \set{0, 0.1, \ldots, 1}$, and $\gamma \in \set{10^{-7}, 10^{-6}, \ldots, 10^{-3}}$. If the optimal combination contains boundaries, then we adjust the search range to make sure the previous optimal hyper-parameter value is within the search range. We iterate this procedure until the optimal hyper-parameters found are inside the search ranges.
We report the value of the hyper-parameters we pick in Table \ref{table:hyper}.

\begin{table}[!htbp]
	\caption{Best hyper-parameters}
	\begin{center}
		\begin{tabular}{l | c | c | c | c}
			\hline
			Region & Model & $\svrCost$ & $\svrEpsilon$ & $\gamma$ \\
			\hline
			Region 1 & Linear-Kernel SVR (4 features) & 0.0625 & 0.3 & \textemdash \\
			Region 1 & Radial-Kernel SVR (4 features) & $2^{12}$ & 0.3 & $10^{-5}$ \\
			Region 1 & Linear-Kernel SVR (3 features) & $2^{14}$ & 0.5 & \textemdash \\
			Region 1 & Radial-Kernel SVR (3 features) & $2^{10}$ & 0.5 & $10^{-3}$ \\ \hline
			Region 2 & Linear-Kernel SVR (4 features) & $2^8$ &  0.1 & \textemdash\\
			Region 2 & Radial-Kernel SVR (4 features) & $2^{16}$ &  0.3 & $10^{-4}$ \\
			Region 2 & Linear-Kernel SVR (3 features) & $2^6$ &  0.1 & \textemdash \\
			Region 2 & Radial-Kernel SVR (3 features) & $2^{10}$ &  0.2 & $10^{-3}$ \\
			\hline
		\end{tabular}
	\end{center}
	\label{table:hyper}
\end{table}

We now turn to evaluating the impact of including $\weightMatrix \bpDemand$ as an input variable in the models tested. As described below, we find that in the region exhibiting spatial autocorrelation, it's beneficial to have the spatial weighted average base-product sales as one of the features. 

Table~\ref{table:pred1} and~\ref{table:pred2} reports the \emph{root mean squared error} (RMSE) and the \emph{mean absolute percent error} (MAPE) based on a 50-time 10-fold cross validation for Region 1 and Region 2, respectively.  For Region 1, which exhibits no spatial autocorrelation, models using three features are significantly better than those with four features. Radial-kernel SVR models slightly outperform the other two types of models. The out-of-sample MAPE of this radial-kernel SVR model is 22.9\%, which is lower than those in \cite{fisher2016optimal}. For Region 2, which has spatial autocorrelation, models using four features are slightly better than those with three features. Similar to Region 1, radial-kernel SVR models outperform the rest two. Also, the out-of-sample MAPE of this model is 16.8\%, which is lower than that those in \cite{fisher2016optimal}. We therefore conclude that radial-kernel SVR models are best for both regions, and that including $\weightMatrix \bpDemand$ for Region 2 increases predictive accuracy.

\begin{table}[h!]
	\caption{Predictive Models Performance of Region 1}
	\begin{center}
		\begin{tabular}{l | c | c | c}
			\hline
			Region 1 & Linear-Kernel SVR & Radial-Kernel SVR & Linear Regression \\
			\hline
			Mean RMSE (4 Features)  & 751.40  & 748.99 & 835.75 \\
			SD RMSE (4 Features)    & 14.73 &16.20 &18.85 \\
			Mean MAPE (4 Features)  & 24.51\%  & 24.19\% & 23.35\% \\
			SD MAPE (4 Features)    & 0.80\%  & 0.91\% & 1.08\% \\
			\hline
			Mean RMSE (3 Features)  & 719.99 & 718.78 & 733.80 \\
			SD RMSE (3 Features)    & 9.28  & 9.61 & 10.36 \\
			Mean MAPE (3 Features)  &23.33\% & 22.90\% & 23.67\% \\
			SD MAPE (3 Features)    &0.29\% & 0.34\% & 0.40\% \\
			\hline
		\end{tabular}
	\end{center}
	\label{table:pred1}
\end{table}

\begin{table}[h!]
	\caption{Predictive Models Performance of Region 2}
	\begin{center}
		\begin{tabular}{l | c | c | c}
			\hline
			Region 2 & Linear-Kernel SVR & Radial-Kernel SVR & Linear Regression \\
			\hline
			Mean RMSE (4 Features)  & 771.38 & 756.47 & 788.43 \\
			SD RMSE (4 Features)   &4.82 & 8.40 & 8.56 \\
			Mean MAPE (4 Features)  &17.06\% &16.84\% &18.02\% \\
			SD MAPE (4 Features)     &0.31\%  &0.27\% &0.40\% \\ \hline
			Mean RMSE (3 Features)  &772.75 &762.03 &791.83 \\
			SD RMSE (3 Features)    &2.82  &6.40 &6.60 \\
			Mean MAPE (3 Features)  &16.37\% &16.26\% &17.92\% \\
			SD MAPE (3 Features)   &0.16\%  &0.22\% &0.25\% \\
			\hline
		\end{tabular}
	\end{center}
	\label{table:pred2}
\end{table}

In the optimization models we use for determining expansion decisions, we will fix the best predictive models for each region and find expansion sites that will maximize expected sales.  The important insight from the analysis conducted in this section is that in the presence of spatial autocorrelation, including the weighted matrix is critical for good predictive performance.

\section{Expansion Optimization}
\label{sec:ExpansionOptimization}

Equipped with a high-performing predictive model, we now turn to determining the optimal set of expansion sites, to maximize expected sales for add-on products given the set of candidate locations. The form of the optimization model depends entirely on the choice of predictive models.  Simpler predictive models result in tractable optimization problems, and more complex models result in highly nonlinear optimization models that require heuristics. 

As described in Section~\ref{sec:ProblemFormulation}, the optimization problem, in its most general form, is~(\ref{eqn:optModel}). This section first explores how this optimization model is specified for the various predictive models employed, and then presents a general purpose greedy heuristic for finding expansion decisions.  We then describe how the solutions we obtain perform in practice on the data made available from our industry partners. 

\subsection{Models with No Spatial Effects}
\label{sec:LinModelNoSpatialEffects}

Consider a predictive model $\hat{f}$ taking the form
\begin{equation}
\tag{LM-NS}
\label{eqn:lm-ns}
\hat{f}(\tilde{\allSites}) = \sum_{i \in \tilde{\allSites}} \linearCost_i .
\end{equation}
This results from, for example, if a linear regression model or a linear-kernel SVR model is employed.  In the former, $\linearCost_i = \beta_1 \bpDemand_i + \beta_3 \householdIncome_i + \beta_4 \population_i + \beta_0$.  In the latter, $\linearCost_i = \sum_{k=1}^{\numberSupportVector} \svrCost_k (\supportVector_{k,1} \bpDemand_i + \supportVector_{k,3} \householdIncome_i + \supportVector_{k,4} \population_i) + b_0$, where $\svrCost_k$ and $\supportVector_{k,j}$ ($j = 1, 2, \textnormal{ and } 4$ in our example) are the standard parameters in SVR models, and $\numberSupportVector$ is the number of support vectors of a specific SVR model.  In either event,~(\ref{eqn:optModel}) reduces to
\begin{align*}
& \max && \sum_{i \in \allSites} \linearCost_i x_i \\
& \textnormal{s.t. } && \sum_{i \in \candidateSizes} x_i = \expansionNumSites \\
&&&
x_i = 1,
&&
\forall i \in \currentSites \\
&&&
x_i \in \set{0,1} 
&&
\forall i \in \candidateSizes.
\end{align*}

Notice that this expansion optimization problem does not involve any association with existing sites that the add-on product is sold at.  Furthermore, this can be solved by simply sorting the predicted add-on-product sales at the candidate sites in a non-increasing order of $\linearCost_i$, and choosing the first $\expansionNumSites$ sites to expand to.  

\subsection{Linear Models with Spatial Effects}

Suppose we incorporated the weight matrix $\weightMatrix$ in a linear regression or linear-kernel SVR. 
Given a collection of sites $\tilde{\allSites}$ offering the add-on product, the model will predict
\begin{equation}
\tag{LM-WS}
\label{eqn:lm-ws}
\hat{f}(\tilde{\allSites}) = \sum_{i \in \tilde{\allSites}} \linearCost_i x_i + \sum_{i \in \tilde{\allSites}} \sum_{j \in \tilde{\allSites} j \neq i} e_{i,j} x_i x_j
\end{equation}
as the total sales of the add-on product.  For a linear regression model, we have
\[
\linearCost_i = \beta_1 \bpDemand_i + \beta_3 \householdIncome_i + \beta_4 \population_i + \beta_0, e_{i,j} = \beta_2 \sum_{j \neq i} \frac{1}{\distance_{i,j}} \bpDemand_j,
\]
and for a linear-kernel SVR, we have
\[
\linearCost_i = \sum_{k=1}^{\numberSupportVector} \svrCost_k (\supportVector_{k,1} \bpDemand_i + \supportVector_{k,3} \householdIncome_i + \supportVector_{k,4} \population_i) + b_0, e_{i,j} = \sum_{k=1}^{\numberSupportVector} \svrCost_k (\supportVector_{k,2} (\sum_{j \neq i} \frac{1}{\distance_{i,j}} \bpDemand_j )).
\]
The expansion optimization model can then be written as  
\begin{align*}
& \max &&
\sum_{i \in \allSites} \linearCost_i x_i  + \sum_{i \in \allSites} \sum_{j \in \allSites, j \neq i} e_{i,j} x_i x_j \\
& \textnormal{s.t. } &&
\sum_{i \in \candidateSizes} x_i = \expansionNumSites \\
&&&
x_i = 1,
&&
\forall i \in \currentSites \\
&&&
x_i \in \set{0,1} 
&&
\forall i \in \candidateSizes.
\end{align*}
The ensuing model is, therefore, a \emph{cardinality-constrained binary quadratic} optimization problem \citep{bertsimas2009algorithm}, which has large literature containing dedicated methods.  For this paper and the data set we use, the optimization model was easily solved by a commercial integer programming solver (\texttt{GUROBI 7.5.1}). 

\subsection{More General Models}

Nonlinear predictive models often result in superior predictive power, as in the case for Region 2.  The resulting optimization models are more complicated.  For example, consider a radial-kernel SVR model with a spatial weight matrix as a part of a feature.  Given a collection of sites $\tilde{N}$, the function form is more complicated than the linear or quadratic form above. We, therefore, devise a heuristic to solve the ensuing optimization model. 


Given any function $\hat{f}(\tilde{\allSites})$ and a number of expansion sites $\expansionNumSites$, one can devise a host of heuristics to plan expansion. We suggest the following, simple heuristic, that can apply to any analytical function.  

Start with no candidate sites, and let $\tilde{\allSites} = \currentSites$.  
\tengcolor{Among all sites in $\candidateSizes$, choose the site $\Phi$ that, if added to $\tilde{\allSites}$, the increase from $\hat{f}(\tilde{\allSites})$ to $\hat{f}(\set{\tilde{\allSites}, \Phi})$ is the most. Then, remove $\Phi$ from \candidateSizes, and add it to $\tilde{\allSites}$. Continue  until $\card{\tilde{\allSites}} - \card{\currentSites} = \expansionNumSites$.}

This heuristic does not guarantee an optimal selection of sites. It will guarantee an optimal selection when employing models with no spatial weight matrix.  Also, in our experimental results, this heuristic was able to match the optimal expansions for all linear models with spatial weight matrix. There is no guarantee that this will happen for other data sets, but given its performance, we use this simple heuristic for the more complicated nonlinear radial-kernel SVR model as well. 

\subsection{Analysis of Results}

In order evaluate the performance of the expansion optimization framework, we compare with a standard practice that an add-on retailer might employ, which determines expansion by choosing those sites for which the base-product sales are highest.  We refer to this expansion as the \emph{baseline} algorithm, or \textbf{BM}.  For any predictive model $P$, we let \textbf{EO-P} be the expansion optimization algorithm using predictive model $P$.

 In order to simulate base-product sales in candidate locations, for each $i \in \candidateSizes$, we draw, independently, $\bpDemand_i$ from a normal distribution with mean $\mu$ and standard deviation $\sigma$.  $\mu$ is taken as the mean of base-product sales in $\currentSites$, and we test for $\sigma = 4^s$, \tengcolor{for $s \in \set{2, 4, 6}$}, in order to evaluate differences in performance for different base-product sales distributions. If $\bpDemand_i$ is drawn to be below the minimum base-product sales in the region, we set $\bpDemand_i$ equal to that minimum.   We simulate this demand ten times per $\sigma$ for the following experiments.  We test for $\expansionNumSites \in \set{1, 2, \ldots, 20}$. 

 We first compare, for the best predictive model in each region, $P^*$, what the average predicted total sales of add-on product will be, using both \textbf{BM} and \textbf{EO-P$^*$}, for each $\expansionNumSites$ tested.   $P^*$ represents radial-kernel SVR models for both regions, with $\weightMatrix \bpDemand$ included as an independent variables in Region 2.  
 
 Table~\ref{tab:expansionResults} report results comparing the expansion decisions.  Each row corresponds to varying $\expansionNumSites$.  The columns report the average increase in additional sales that result from employing \textbf{EO-P$^*$} over \textbf{BM}. In particular, let $z^0$ be the total sales of add-on product estimated by $P^*$ without any expansion.  For any $\expansionNumSites$, let $z_e^\expansionNumSites$ be the total add-on sales predicted using \textbf{EO-P$^*$} and $z_b^\expansionNumSites$ be the total add-on sales predicted using \textbf{BM}.  The table reports $\frac{z_e^\expansionNumSites - z_b^\expansionNumSites}{z_b^\expansionNumSites - z^0} \cdot 100 \%$.  The higher this value, the more gains that are realized by using the framework we propose. 

\begin{table}[t!]
	\caption{Average percent increase in additional sales by using \textbf{EO-P$^*$} over \textbf{BM} in both regions for various base-product demand distributions}
	\label{tab:expansionResults}
	\begin{center}
		\begin{tabular}{  l | l l l | l l l } \hline
			&		\multicolumn{3}{c |}{Region 1} &
			\multicolumn{3}{c }{Region 2} \\ \hline
			$\expansionNumSites$ & $s = 2$ & $s = 4$& $s = 6$ & $s = 2$& $s = 4$ & $s = 6$  \\ \hline
1  & 5.48\% & 4.45\% & 2.65\% & 3.18\% & 3.45\% & 2.11\% \\
2  & 4.58\% & 4.39\% & 2.65\% & 4.43\% & 4.21\% & 2.26\% \\
3  & 4.64\% & 4.22\% & 2.47\% & 5.50\% & 4.54\% & 2.43\% \\
4  & 4.52\% & 4.03\% & 2.25\% & 5.63\% & 5.64\% & 2.67\% \\
5  & 4.59\% & 4.07\% & 2.17\% & 5.51\% & 5.54\% & 2.54\% \\
6  & 4.42\% & 3.95\% & 2.07\% & 5.19\% & 5.57\% & 2.64\% \\
7  & 4.30\% & 3.78\% & 1.93\% & 4.88\% & 5.98\% & 2.58\% \\
8  & 4.21\% & 3.66\% & 1.84\% & 4.69\% & 5.67\% & 2.43\% \\
9  & 4.18\% & 3.65\% & 1.77\% & 4.55\% & 5.46\% & 2.62\% \\
10 & 4.00\% & 3.57\% & 1.75\% & 4.44\% & 5.32\% & 2.78\% \\
11 & 3.89\% & 3.51\% & 1.70\% & 4.38\% & 5.10\% & 3.03\% \\
12 & 3.85\% & 3.49\% & 1.63\% & 4.50\% & 5.08\% & 2.96\% \\
13 & 3.79\% & 3.44\% & 1.59\% & 4.68\% & 4.96\% & 2.84\% \\
14 & 3.72\% & 3.43\% & 1.51\% & 4.53\% & 4.79\% & 2.86\% \\
15 & 3.61\% & 3.34\% & 1.52\% & 4.39\% & 4.60\% & 2.96\% \\
16 & 3.54\% & 3.29\% & 1.51\% & 4.28\% & 4.45\% & 2.95\% \\
17 & 3.49\% & 3.21\% & 1.48\% & 4.27\% & 4.35\% & 3.02\% \\
18 & 3.44\% & 3.17\% & 1.44\% & 4.22\% & 4.43\% & 2.91\% \\
19 & 3.41\% & 3.13\% & 1.43\% & 4.33\% & 4.54\% & 2.95\% \\
20 & 3.38\% & 3.10\% & 1.39\% & 4.62\% & 4.64\% & 2.94\% \\ \hline
\end{tabular}
\end{center}
\end{table}

\subsection{Managerial Insights}

Our results provide clear evidence that making expansion decisions for add-on products should not be done without more readily understanding how sales will be affected by the expansion decisions made.  A standard approach to expanding by only considering base-product sales can lead to far worse expansion decisions than the more complex but effective predictive and optimization framework developed in this paper. 

As shown in Table~\ref{tab:expansionResults}, when $\expansionNumSites$ is small, we realize larger gains when there is no spatial autocorrelation (Region 1). However, this switches when $\expansionNumSites$ increases. This indicates that our suggested approach is particularly useful when the retailer is considering a large expansion strategy in the presence of spatial autocorrelation. Also, in general, gains decrease with $\expansionNumSites$ when there is no spatial autocorrelation (Region 1), and increase with $\expansionNumSites$ where there is (Region 2).

Table~\ref{tab:expansionResults} exhibits the gains the add-on product seller can realize by employing our strategy.  When comparing with the results obtained with \textbf{BM}, the add-on product seller can realize additional sales of over 5\%, simply by choosing better expansion decision making.  This depends, however, on the distribution of the demand for base-products.  For example, if $\expansionNumSites = 10$, in Region 1, when the variance is low ($s=2$), we expect to achieve $4.00\%$ increase in sales simply by choosing a better set of locations.  As the variance increases, the difference between \textbf{BM} and \textbf{EO-P$^*$} becomes smaller. The same trend appears in Region 2 as well. 

\tengcolor{This trend} appears to be more extreme when base-product sales at candidate sites do not vary much among the candidate locations.  In such a case, relying on base-product sales does not provide enough information. \tengcolor{Hence, the information from the neighborhood become more important.} Even when the variation does exist among candidate sites, using simple expansion model relying only on base-product sales will not achieve optimal results.  In order to capture complex relationships between demographic information and the interplay between the demand from neighboring base-product retail sites, predictive models using spatial weight matrices can improve prediction power, thereby allowing optimization models to automate the selection of desirable expansion sites. 

As far as the net increase in sales, if $\expansionNumSites = 20$ and $s = 2$, the difference in additional sales of add-on product in Region 1 is 3429, and in Region 2 it is 3678.  For a tight-margin business, this may have significant impacts on its bottom line, particularly because this increase in sales does not require anything more than selecting better expansion sites. 

\begin{figure}[t!] 
\begin{minipage}[b]{0.5\linewidth}
	\centering
		\includegraphics[scale=0.4]{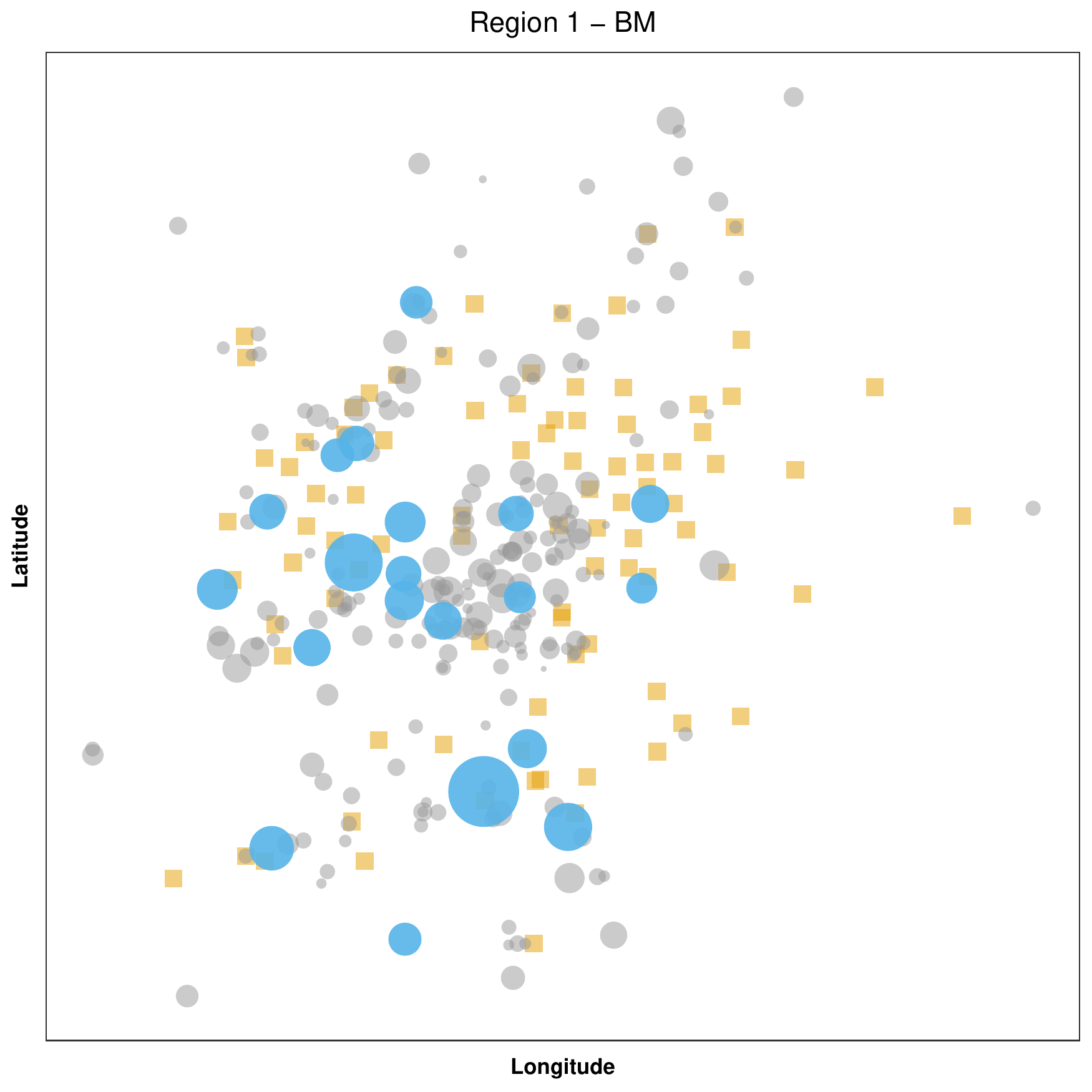} 
	\end{minipage}
	\begin{minipage}[b]{0.5\linewidth}
		\centering
		\includegraphics[scale=0.4]{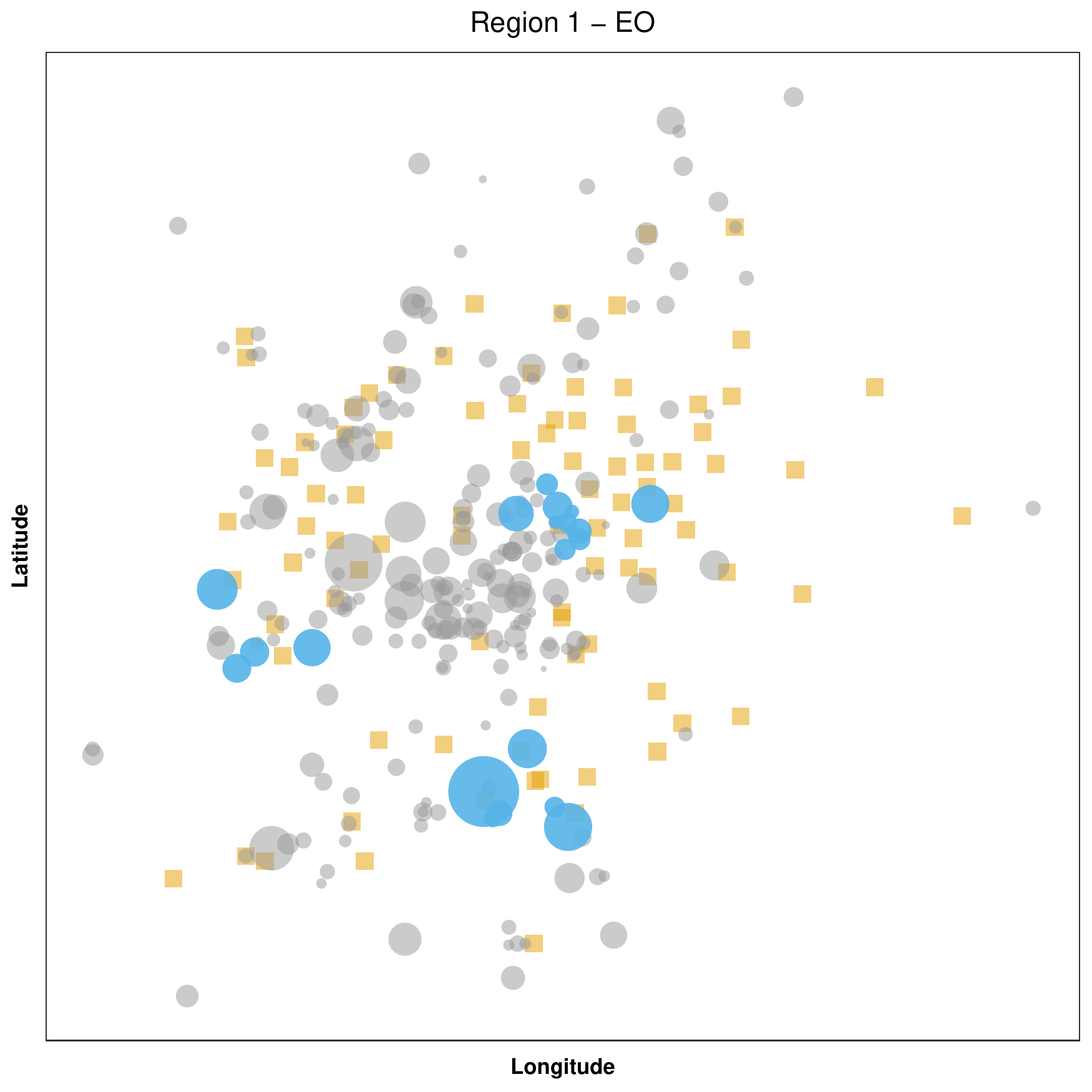} 
	\end{minipage}
	\caption{Best known expansion sites for Region 1 using (left) \textbf{BM} and (right)  \textbf{EO-P$^*$}.  Current sites drawn as yellow squares.  Candidate sites drawn as circles, colored blue if selected and gray otherwise. The relative sizes of the circles correspond to the total base-product sales at the corresponding site.} 
	\label{fig:expansionSitesRegion1} 	
\end{figure}

\begin{figure}[t!] 
\begin{minipage}[b]{0.5\linewidth}
	\centering
		\includegraphics[scale=0.4]{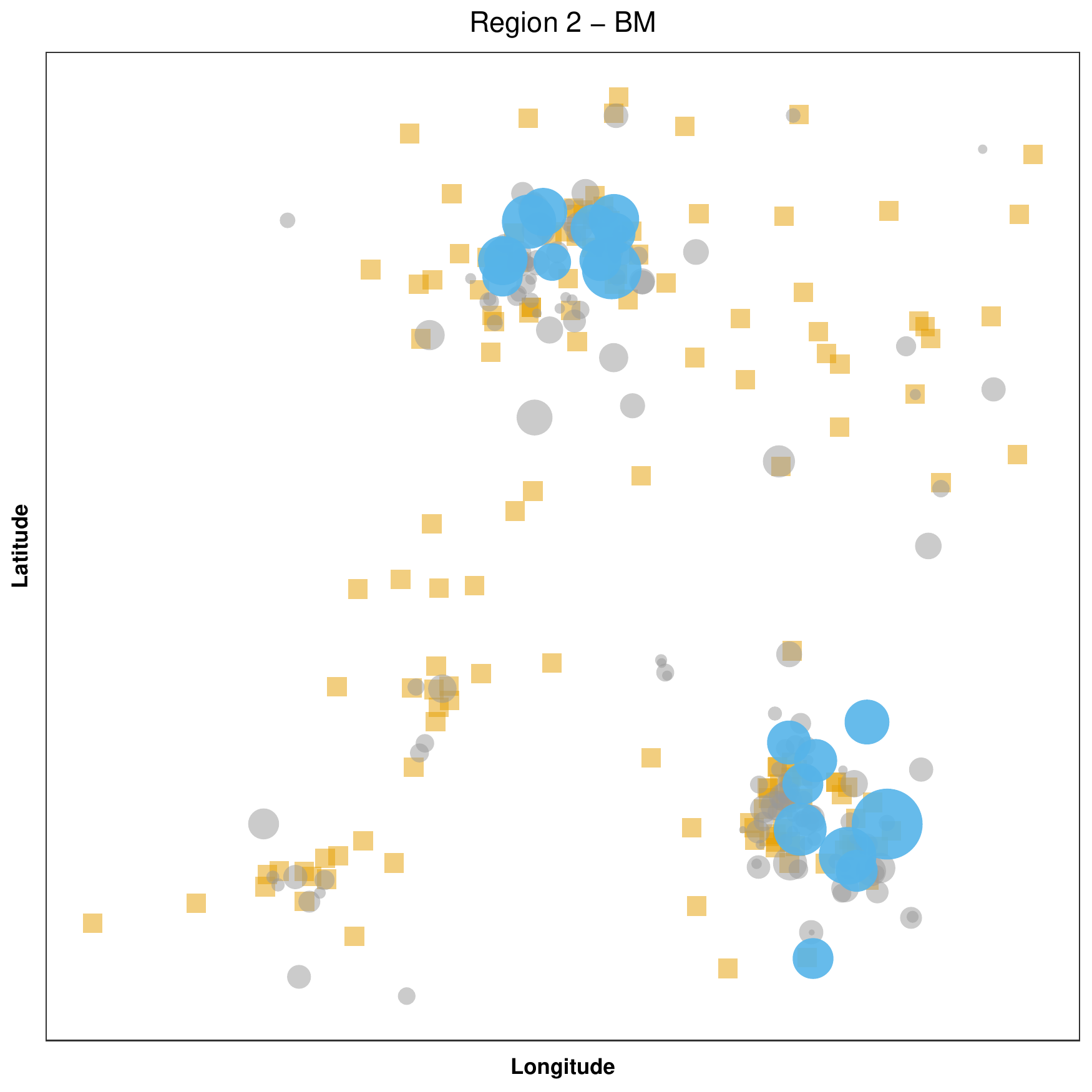} 
	\end{minipage}
	\begin{minipage}[b]{0.5\linewidth}
		\centering
		\includegraphics[scale=0.4]{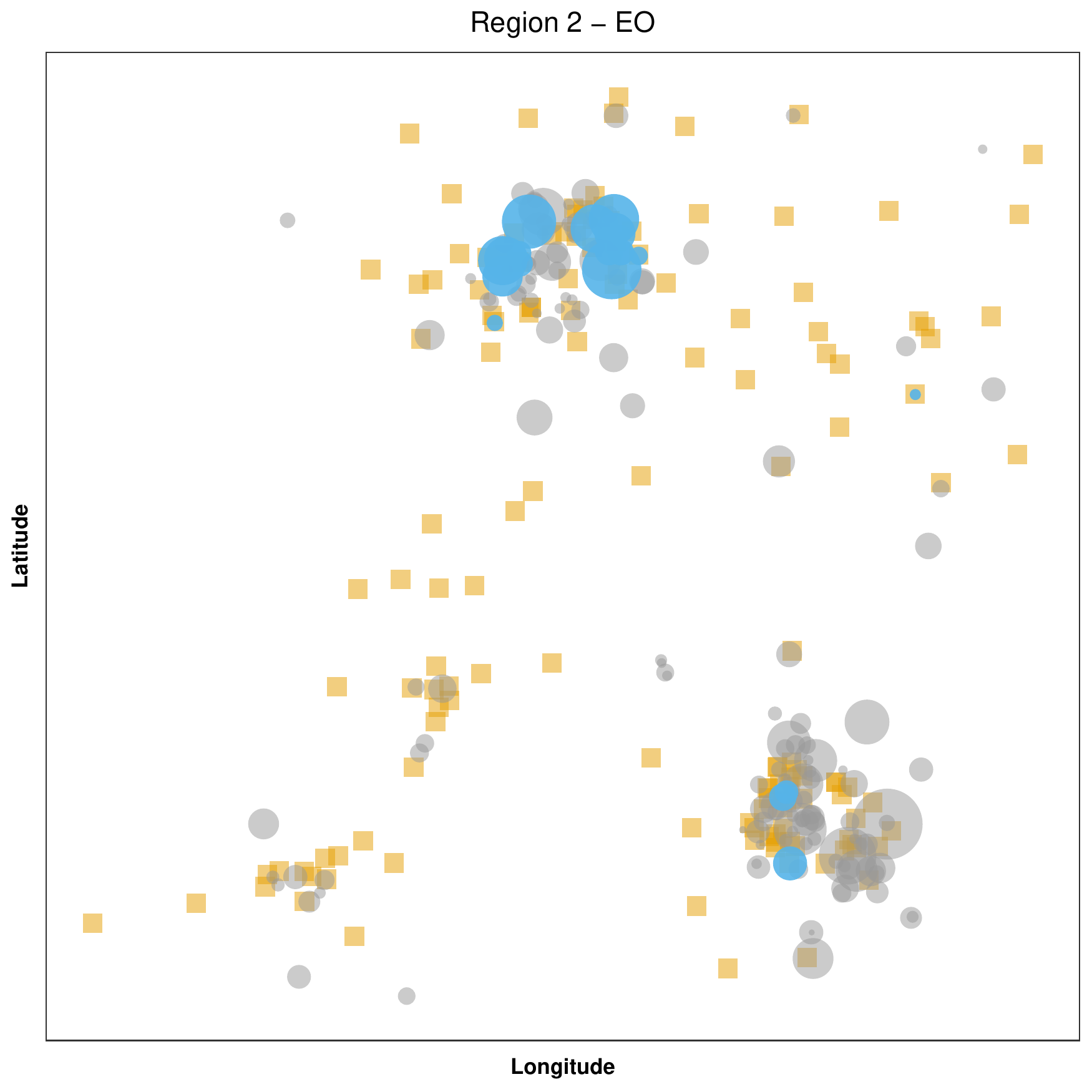} 
	\end{minipage}
	\caption{Best known expansion sites for Region 2 using (left) \textbf{BM} and (right)  \textbf{EO-P$^*$}. Current sites drawn as yellow squares.  Candidate sites drawn as circles, colored blue if selected and gray otherwise.  The relative sizes of the circles correspond to the total base-product sales at the corresponding site. } 
	\label{fig:expansionSitesRegion2} 	
\end{figure}

It is also interesting to view the geographic dispersion and selection of sites, comparing the expansion decisions made using \textbf{BM} with \textbf{EO-P$^*$} for both regions.  This is depicted in Figure~\ref{fig:expansionSitesRegion1} (for Region 1) and Figure~\ref{fig:expansionSitesRegion2} (for Region 2) for one instance generated with $s = 6$.  Each figure displays a point for every site at its geographic location.  The current sites are squares and colored orange, and the candidate sites are circles, colored blue if chosen for expansion and gray otherwise.  Furthermore, the points corresponding to the candidate sites are gradient sized to depict relative base-product sales (higher sales, larger circle).  In both figures, the left plot depicts the solution obtained using \textbf{BM}, and the right plot depicts the solution obtained using \textbf{EO-P$^*$}.

These figures depict more clearly how the expansion decisions differ.  In particular, we see that the expansion decisions made through \textbf{EO-P$^*$} take advantage of local clusters, and although selection leans towards those sites with large base-product sales, it is often better to consider location impact over simply base-product sales.

It is clear that using the optimization models for a fixed predictive model will always result in better results than using the baseline algorithm when evaluated on the predictive model used within the optimization framework.  In order to ensure that the results are robust, we suggest the following procedure.   Take the solutions obtained by \textbf{EO-P$^*$} and those obtained by \textbf{BM}, and compare them using \emph{different} predictive models. Since different predictive models provide estimation using varied structure, it is not clear whether a solution that is superior for one predictive model will also be superior for other predictive models, built using alternative structural assumptions.  If a solution is better across multiple predictive models, this makes the solution even more desirable.

\begin{table}[t!]
	\centering
	\caption{Robustness check for solution obtained by $P^*$ for Region 1}
	\label{tab:RobustCheckRegion1}
	\begin{tabular}{l|lll|lll|lll} \hline
		&
		\multicolumn{3}{c}{$s=2$}
		&
		\multicolumn{3}{c}{$s=4$}
		&
		\multicolumn{3}{c}{$s=6$}
		\\
		\hline
		$\expansionNumSites$ 
		&
		\textbf{LR}
		&
		\textbf{LK}
		&
		\textbf{RK}
		&
		\textbf{LR}
		&
		\textbf{LK}
		&
		\textbf{RK}
		&
		\textbf{LR}
		&
		\textbf{LK}
		&
		\textbf{RK}
		
		\\ 
		\hline
1  & 3.34\% & 4.92\% & 5.48\% & 2.61\% & 3.94\% & 4.45\% & 0.61\% & 1.98\% & 2.65\% \\
2  & 2.89\% & 4.27\% & 4.58\% & 2.69\% & 4.04\% & 4.39\% & 0.90\% & 2.27\% & 2.65\% \\
3  & 2.97\% & 4.39\% & 4.64\% & 2.65\% & 4.03\% & 4.22\% & 1.01\% & 2.21\% & 2.47\% \\
4  & 2.92\% & 4.33\% & 4.52\% & 2.53\% & 3.85\% & 4.03\% & 0.90\% & 2.08\% & 2.25\% \\
5  & 2.98\% & 4.43\% & 4.59\% & 2.56\% & 3.90\% & 4.07\% & 0.85\% & 2.04\% & 2.17\% \\
6  & 2.96\% & 4.43\% & 4.42\% & 2.57\% & 3.94\% & 3.95\% & 0.81\% & 1.96\% & 2.07\% \\
7  & 2.90\% & 4.35\% & 4.30\% & 2.46\% & 3.78\% & 3.78\% & 0.73\% & 1.82\% & 1.93\% \\
8  & 2.86\% & 4.29\% & 4.21\% & 2.41\% & 3.71\% & 3.66\% & 0.72\% & 1.81\% & 1.84\% \\
9  & 2.88\% & 4.33\% & 4.18\% & 2.43\% & 3.75\% & 3.65\% & 0.70\% & 1.76\% & 1.77\% \\
10 & 2.79\% & 4.22\% & 4.00\% & 2.40\% & 3.70\% & 3.57\% & 0.68\% & 1.73\% & 1.75\% \\
11 & 2.71\% & 4.08\% & 3.89\% & 2.37\% & 3.65\% & 3.51\% & 0.66\% & 1.66\% & 1.70\% \\
12 & 2.67\% & 4.02\% & 3.85\% & 2.36\% & 3.64\% & 3.49\% & 0.64\% & 1.61\% & 1.63\% \\
13 & 2.62\% & 3.94\% & 3.79\% & 2.33\% & 3.58\% & 3.44\% & 0.63\% & 1.55\% & 1.59\% \\
14 & 2.56\% & 3.84\% & 3.72\% & 2.31\% & 3.57\% & 3.43\% & 0.58\% & 1.48\% & 1.51\% \\
15 & 2.49\% & 3.75\% & 3.61\% & 2.25\% & 3.46\% & 3.34\% & 0.61\% & 1.49\% & 1.52\% \\
16 & 2.44\% & 3.67\% & 3.54\% & 2.21\% & 3.40\% & 3.29\% & 0.61\% & 1.47\% & 1.51\% \\
17 & 2.40\% & 3.60\% & 3.49\% & 2.16\% & 3.32\% & 3.21\% & 0.60\% & 1.44\% & 1.48\% \\
18 & 2.37\% & 3.55\% & 3.44\% & 2.13\% & 3.26\% & 3.17\% & 0.59\% & 1.43\% & 1.44\% \\
19 & 2.34\% & 3.50\% & 3.41\% & 2.09\% & 3.19\% & 3.13\% & 0.60\% & 1.42\% & 1.43\% \\
20 & 2.31\% & 3.45\% & 3.38\% & 2.06\% & 3.16\% & 3.10\% & 0.57\% & 1.37\% & 1.39\% \\ \hline
	\end{tabular}
\end{table}

\begin{table}[t!]
	\centering
	\caption{Robustness check for solution obtained by $P^*$ for Region 2}
	\label{tab:RobustCheckRegion2}
	\begin{tabular}{l|lll|lll|lll} \hline
		&
		\multicolumn{3}{c}{$s=2$}
		&
		\multicolumn{3}{c}{$s=4$}
		&
		\multicolumn{3}{c}{$s=6$}
		\\
		\hline
		$\expansionNumSites$ 
		&
		\textbf{LR}
		&
		\textbf{LK}
		&
		\textbf{RK}
		&
		\textbf{LR}
		&
		\textbf{LK}
		&
		\textbf{RK}
		&
		\textbf{LR}
		&
		\textbf{LK}
		&
		\textbf{RK}
		
		\\ 
		\hline
1  & 6.07\%  & -2.50\% & 3.18\% & 5.69\%  & -4.68\% & 3.45\% & 4.21\% & -5.40\% & 2.11\% \\
2  & 7.64\%  & 0.47\%  & 4.43\% & 7.32\%  & -0.50\% & 4.21\% & 6.55\% & -4.91\% & 2.26\% \\
3  & 10.85\% & -0.57\% & 5.50\% & 10.69\% & -1.37\% & 4.54\% & 7.82\% & -4.33\% & 2.43\% \\
4  & 10.97\% & -0.22\% & 5.63\% & 11.26\% & -0.38\% & 5.64\% & 9.24\% & -3.42\% & 2.67\% \\
5  & 10.91\% & -0.28\% & 5.51\% & 11.38\% & -0.59\% & 5.54\% & 9.13\% & -2.62\% & 2.54\% \\
6  & 11.48\% & 0.39\%  & 5.19\% & 12.38\% & -0.12\% & 5.57\% & 8.57\% & -2.12\% & 2.64\% \\
7  & 10.65\% & 0.05\%  & 4.88\% & 12.29\% & -0.30\% & 5.98\% & 8.84\% & -1.85\% & 2.58\% \\
8  & 10.28\% & 0.22\%  & 4.69\% & 11.80\% & 0.00\%  & 5.67\% & 8.27\% & -1.78\% & 2.43\% \\
9  & 9.75\%  & 0.43\%  & 4.55\% & 11.59\% & 0.09\%  & 5.46\% & 8.46\% & -1.67\% & 2.62\% \\
10 & 9.92\%  & 0.45\%  & 4.44\% & 11.80\% & 0.02\%  & 5.32\% & 8.63\% & -1.60\% & 2.78\% \\
11 & 10.31\% & 0.36\%  & 4.38\% & 11.95\% & -0.11\% & 5.10\% & 8.31\% & -1.09\% & 3.03\% \\
12 & 13.53\% & 0.32\%  & 4.50\% & 14.18\% & -0.12\% & 5.08\% & 8.00\% & -0.97\% & 2.96\% \\
13 & 13.88\% & 0.32\%  & 4.68\% & 14.50\% & -0.03\% & 4.96\% & 7.77\% & -0.97\% & 2.84\% \\
14 & 13.35\% & 0.51\%  & 4.53\% & 14.27\% & 0.13\%  & 4.79\% & 8.39\% & -0.86\% & 2.86\% \\
15 & 12.82\% & 0.46\%  & 4.39\% & 14.15\% & 0.19\%  & 4.60\% & 8.27\% & -0.81\% & 2.96\% \\
16 & 12.60\% & 0.78\%  & 4.28\% & 14.06\% & 0.40\%  & 4.45\% & 8.00\% & -0.68\% & 2.95\% \\
17 & 13.16\% & 0.69\%  & 4.27\% & 13.94\% & 0.50\%  & 4.35\% & 7.89\% & -0.57\% & 3.02\% \\
18 & 13.34\% & 0.70\%  & 4.22\% & 14.16\% & 0.61\%  & 4.43\% & 7.50\% & -0.54\% & 2.91\% \\
19 & 13.01\% & 0.77\%  & 4.33\% & 14.14\% & 0.73\%  & 4.54\% & 7.36\% & -0.48\% & 2.95\% \\
20 & 13.96\% & 0.82\%  & 4.62\% & 14.25\% & 0.70\%  & 4.64\% & 7.22\% & -0.47\% & 2.94\% \\ \hline
	\end{tabular}
\end{table}

Consider Tables~\ref{tab:RobustCheckRegion1} and~\ref{tab:RobustCheckRegion2}.  For both regions and each $s$, we use the solution from \textbf{EO-P$^*$} and evaluate the number of sales predicted by the other predictive models, to see if the expansion decision remains favorable over \textbf{BM} when evaluated on another predictive model.  This will enable managers to have more confidence in the solutions. 

Let \textbf{LR} be the linear regression model, \textbf{LK} be the linear-kernel SVR model, and \textbf{RK} be the radial-kernel SVR model, where we include $Wg$ as a predictor in the models only for Region 2.   Each entry in the tables corresponds to the percent increase in additional expected sales realized from using the solution obtained by \textbf{EO-P$^*$}  over the solution obtained by \textbf{BM}.  Interestingly, the solution found by \textbf{EO-P$^*$} are almost always better than \textbf{BM} (i.e., most of the numbers in Tables~\ref{tab:RobustCheckRegion1} and~\ref{tab:RobustCheckRegion2} are positive), independent of the predictive model that it is evaluated on.  The exception is when we plug in \textbf{EO-P$^*$} into \textbf{LK} in Region 2.

\section{Conclusion and Future Work}
\label{sec:conclusion}

In this paper, we provide a novel framework for jointly applying predictive modeling and optimization to expansion decisions for add-on products.  Through collaboration with an industrial partner, we use point of sales data to generate predictive models and formulate optimization models over those predictive models to automate expansion decisions. Our results indicate that making expansion decisions using the framework described in this paper can result in substantial increases in expected sales over a baseline algorithm that only takes into account base-product sales.  

This work has myriad possible directions for expansion. First, if given more complete and complex data, one could build more elaborate predictive models that provide a more fine-grained picture of what drives sales for add-on products. For example, a more detailed customer-level data on purchase behavior at different locations can be incorporated to develop improved predictive models. This can also drive the prescriptive model, for instance,  expanding to a new site frequented by existing customers may increase convenience for them but may limit market expansion to access new customers. Considering such nuances in customer behavior can lead to better location choice models. Second, these results should be replicated to more settings where expansion is limited by geographic constraints.  Third, the dedicated solution methodology for solving optimization models resulting from non-linear predictive models should be investigated, so that expansion decisions are made optimally as opposed to by heuristics.  Although the heuristic performs well for the company that we partnered with, there is no proof of optimality, and it is possible, and probable, that better collections of expansion sites exist. 

\clearpage

\bibliographystyle{informs2014} 
\bibliography{ref} 

\end{document}